\documentstyle[prl,aps,epsf]{revtex}\def\narrowtext{}\tighten\twocolumn %camera-ready
\begin{document}
\draft

%\title{Comment on ''Phase Diagram of La$_{2-x}$Sr$_x$CuO$_4$ Probed in the Infrared:
% Imprints of Charge Stripe Excitations''}

%\author{
%S.~Tajima$^{1)}$ and S.~Uchida$^{2)}$ }

%\address{
%1)Superconductivity Research Laboratory-ISTEC, Tokyo 135-0082, Japan\\
%2)Dept. of Physics, The University of Tokyo, Tokyo 113-8656, Japan\\
%}

%\date{\today}
%\maketitle  %preprint

%\address{\begin{minipage}[t]{6.0in}  %camera-ready
%\begin{abstract}%

%\typeout{polish abstract}
%\end{abstract}
%\pacs{}
%\end{minipage}}  %camera-ready  

%\maketitle  %camera-ready
\narrowtext

{\it {\bf Comment on ''Phase Diagram of La$_{2-x}$Sr$_x$CuO$_4$ Probed in the Infrared:
Imprints of Charge Stripe Excitations''}}

Recently Lucarelli {\it et al.} have reported\cite{lucarelli} 
temperature-dependence of the in-plane 
optical reflectivity of La$_{2-x}$Sr$_x$CuO$_4$ over a wide doping range,  
focusing on the infrared peaks at 30 cm$^{-1}$ (for $x$=0.12), 250 cm$^{-1}$ and 510 cm$^{-1}$.  
They interpreted the first peak (30 cm$^{-1}$) as a signature of 
charge stripe ordering, while the latter two (250 cm$^{-1}$ and 
510 cm$^{-1}$) are attributed to the polaronic charge excitations.  
However, careful readers would notice that the reported spectra are largely 
different from those so far measured on the same system. 
As we illustrate below, 
all these peaks are caused by an uncontrolled leakage of the
$c$-axis reflectivity into the measured spectra. 

First, we show that the absorption peaks at 250 cm$^{-1}$ and 510 cm$^{-1}$ 
are nothing but the $c$-axis phonon modes (A$_{2u}$).  
The reported TO-phonon frequencies for two of the three 
A$_{2u}$-modes\cite{collins}
coincide well with the above two frequencies.  
In Fig.1 we compare the inset spectra of Fig.1c in ref.[1] with the 
purely $c$-polarized spectrum of La$_{1.96}$Sr$_{0.04}$CuO$_4$. 
This clearly demonstrates that Lucarelli {\it et al.} observed  the spectra 
mixed with the $c$-axis component.   
A similar mixing is more or less observed in most of their samples except 
for x=0 and 0.26.

Second, the $c$-component mixing seriously affects the reflectivity values 
below $\sim$200 cm$^{-1}$. 
At low tempreatures, 
the reflectivity is close to unity for $E \parallel ab$, 
while it decreases with lowering $\omega$ to less than 0.5 
for $E \parallel c$, 
reflecting the incoherent charge dynamics in the $c$-axis direction. 
If a $c$-component is mixed into the measured in-plane spectrum,
then the measured reflectivity tends to decrease with decreasing $\omega$,
creating an artificial new absorption peak in the Kramers-Kronig 
transformed conductivity spectrum.  
Note that the spectra reported 
by the other groups\cite{gao,dumm} do not show such a pronounced peak 
below 100 cm$^{-1}$. 
Although Lucarelli {\it et al.} mentioned 
that their spectrum for $x$=0.12 is consistent with the result
by Dumm {\it et al.}\cite{dumm}, 
the gigantic peak at 30 cm$^{-1}$ is not seen in the latter.

Finally, we point out a non-systematic doping dependence 
of reflectivity spectrum 
seen in Fig.1 of ref.[1].
For example, the 510 cm$^{-1}$ peak that is substantially weakened at $x$=0.15
develops again for x=0.19,
which strongly suggests an accidental mixing of $c$-component\cite{basov}.
As to the origin of $c$-component mixing,
there are several possibilities
such as polarizer leakage, mis-cutting and multi-domain structure of crystals
grown by traveling-solvent-floating-zone method.
From the non-systematic spectral change with $x$,
it is speculated that the source of $c$-component and the mixing rate may be
different from sample to sample in ref.[1].
It should be noted that 
even a small amount of admixture of the $c$-component seriously affects 
the optical spectrum, 
whereas neutron scattering and/or transport measurements are more robust
against a few percent mixture of different crystal angles.

In summary, the three infrared peaks observed in ref.[1] in addition to the
in-plane phonon peaks do most 
certainly originate from the $c$-axis component mixed into the in-plane 
spectra.  
Therefore, neither the $d$-band scenario nor the charge stripe dynamics can be 
deduced from these experimental results.\\
\\
S. Tajima$^{1)}$, S. Uchida$^{2)}$, D. van der Marel$^{3)}$
and D. N. Basov$^{4)}$\\
\\
$^{1)}$Superconductivity Research Laboratory, ISTEC, Tokyo 135-0062, Japan\\
$^{2)}$Dept. of Physics, The University of Tokyo, Tokyo 113-8656, Japan\\
$^{3)}$Material Science Center, University of Groningen,
Nijenborgh 4, 9747AG, Groningen, The Netherland\\
$^{4)}$Dept. of Physics, University of California at San Diego,
La Jolla, California 92093-0319

\begin{figure}
\epsfxsize=3.2in
\epsfbox{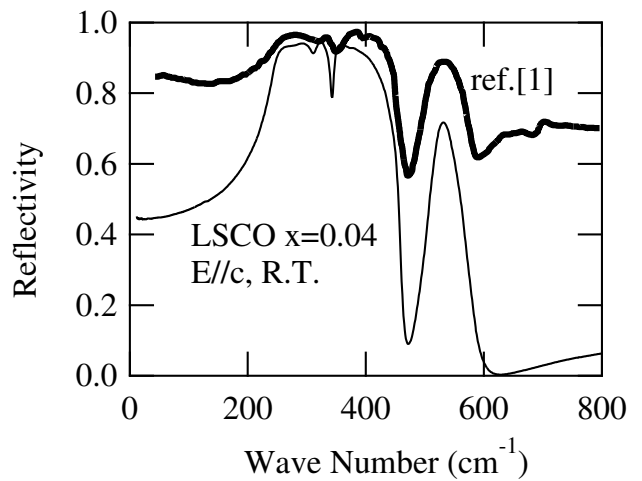}
\vspace{0.5cm}
\caption{Comparison of the reflectivity spectrum of 
La$_{1.95}$Sr$_{0.05}$CuO$_4$ (thick curve) from
Fig.1c inset of ref.[1] with the $c$-polarized spectrum 
of La$_{1.96}$Sr$_{0.04}$CuO$_4$ (thin curve)[7]}
\label{fig1}
\end{figure}


\begin{references}
\bibitem{basov}Also, the magnitude of mid-infrared conductivity 
for $x$=0.03 exceeds that of
both $x$=0.05 and 0.07,
in stark contrast with the established doping dependence[6].
\end{references}
\end{document}